\documentstyle[sprocl]{article}

\def\be{\begin{equation}}
\def\ee{\end{equation}}
\def\bea{\begin{eqnarray}}
\def\eea{\end{eqnarray}}
\begin{document}

\title{WIGNER TRANSFORM AND QUANTUM-LIKE CORRECTIONS FOR 
CHARGED-PARTICLE BEAM TRANSPORT}

\author{ R. FEDELE ~and~ V. I. MAN'KO~\footnote{Permanent address: P.N.
Lebedev Physical Institute, Moscow, Russia}}

\address{Dip. di Scienze Fisiche, Universit\'a
``Federico II'' and INFN, Napoli, Italy}


\maketitle\abstracts{It is shown that
charged-particle beam transport in the paraxial approximation can be
effectively described with a quantum-like picture in {\it
semiclassical approximation}. In particular,
the classical Liouville equation can be suitably replaced by a
von Neuman-like equation. 
Relevant remarks concerning 
the standard classical description of the beam transport are
given.}

It is well known that {\it thermal spreading among the electronic 
rays} is a typical effect that takes place in 
charged-particle beam transport in free space.\cite{lawson}
In 2-D case, we denote with $x$ and $z$ the transverse and beam 
propagation coordinates, respectively. By using a statistical 
description, one can introduce with the second-order moments:
$\sigma_x (z)=\langle x^2
\rangle^{1/2}$, 
$\sigma_p (z)=\langle p^2\rangle^{1/2}$, 
and $\sigma_{xp}=\langle xp\rangle$ \,($p\equiv dx/dz$ being a 
dimensionless single-particle linear momentum conjugate of $x$, where 
$z$ plays the role of a time-like variable)
the following diffusion coefficient called the emittance,\cite{lawson}
$\epsilon =2\left[\langle x^2 
\rangle\langle p^2 \rangle-\langle xp
\rangle^2\right]^{1/2},$ which for
linear lens and in free space is an invariant.
From this expression, in particular, we have
$\sigma _x\sigma _p\geq \epsilon /2$, which represents a sort of
{\it uncertainty relation} even if the particle beam is a {\it classical
system}. It is easy to prove that~\cite{lawson}
$(\epsilon /2)=v_{\rm th}\sigma_0/c$, where
$v_{\rm th}$ is the {\it thermal velocity} of the
system (we assume $v_{\rm th}/c \ll 1$), 
and $\sigma_0\equiv\langle x^2\rangle_{z=0}^{1/2}$.

Thus, for finite temperature the determination of an
electronic ray at the arbitrary $x$-position of the transverse plane 
given at each $z$ is affected by an intrinsic uncertainty that cannot 
be reduced to zero.
For a finite emittance, we need to assign a
probability, say $P_{x} (x,z)$, 
(in principle, positive and finite) of finding an electronic ray at
the transverse location $x$ in the plane for given $z$.
To this end, we introduce the phase-space density
distribution $\rho (\overline{x},p,\overline{z})$, which obeys to the 
following classical Liouville equation for the
electronic rays,
$${\partial \rho \over \partial \overline{z}}+p{\partial
\rho\over\partial \overline{x}
}-\left({\partial \overline{U}\over\partial \overline{x}
}\right){\partial\rho\over\partial p}=0,\quad
\overline{z}\equiv{z\over 2\sigma_0}\,,\quad \overline{x}
\equiv{x\over 2\sigma_0}\,,$$ 
$\overline{U}=
\overline{U}(\overline{x},\overline{z})$ being an effective potential 
acting on the system.
Since for finite emittance the indistinguishability among two or more
rays due to the thermal spreading is of the order of $\eta\equiv \epsilon
/2\sigma_0=v_{\rm th}/c \ll 1$, \,$\partial
\overline{U}/\partial \overline{x}$
can be conveniently
replaced by a symmetrized Schwarz-like finite difference
ratio
$${\partial \overline{U}\over\partial x}{\partial\over \partial p }\approx
{\overline{U}(\overline{x}+\eta /2)-\overline{U}(\overline{x}-\eta /2)\over
\eta}{\partial\over \partial p }\,.$$
This {\it transition} based on physical arguments is
partially a change of partial differential equation
to a differential-difference equation,
which may be considered as
ansatz of a {\it deformation} of the Liouville equation.
Furthermore,
given the smallness of $\eta$, multiplying by the imaginary
unit $i$ both numerator and denominator of the last term of the
l.h.s., we have
$${\overline{U}(\overline{x}+\eta /2)-\overline{U}(\overline{x}-\eta /2)
\over i\eta}~i{\partial \over \partial p}\approx \frac{1}{i\eta}
\left[\overline{U}\left(\overline{x}+{i\eta\over 2}{\partial\over\partial
p}\right)-\overline{U}\left(\overline{x}-{i\eta\over
2}{\partial\over\partial p}\right)\right].$$
Thus, going back to the old variables $x$ and $z$,
it finally results that the classical Liouville equation is {\it 
deformed} in the following von Neumann equation
\begin{equation}\label{17bis} 
\left\{{\partial \over \partial z} + p {\partial \over \partial x}
+{i \over \epsilon} \left[ U\left(x + i {\epsilon \over 2}
{\partial \over \partial p}\right)-  U\left(x - i {\epsilon \over 2}
{\partial \over \partial p}\right)\right] \right\} \rho_w=0\,,
\end{equation}
where the {\it deformed} distribution function 
$\rho_{w}(x,p,z)$ is a sort of Wigner-like function.\cite{wigner}
It is obvious, that Eq.~(\ref{17bis}) has the form of a quantum-like 
phase-space equation for electronic rays, where $\hbar$ and the time
$t$ are replaced by the emittance $\epsilon$
and the propagation coordinate $z$, respectively.
However, some aspects have to be discussed.
$$(\mbox {i}).~\mbox{Since}\quad 
\overline{U}(\overline{x}+{i\eta\over 2}{\partial\over\partial
p})-\overline{U}(\overline{x}-{i\eta\over 2}{\partial\over\partial
p})={\partial\overline{U}\over\partial\overline{x}}~
i\eta{\partial\over\partial p}+O\left(\eta^3{\partial^3\over\partial
p^3}\right),$$ 
the above approximation is equivalent to assume that terms
$O\left(\eta^3\,\partial^3/\partial p^3\right)$ 
are small corrections compared to the
lower-order ones, according to the paraxial approximation.
Consequently, from the quantum-like point of view, approximation
obtained by the above deformation plays the role analogous to the one played
by {\it semiclassical approximation}.\cite{heller}

(ii).~While $\rho (x,p,z)$ is introduced in a classical framework
and it is positive definite, the function $\rho_w (x,p,z)$ is
introduced in a quantum-like framework, which plays the role of an
{\it effective} description taking into account the thermal spreading
among the electronic rays. In this context, $\rho_w
(x,p,z)$ cannot be used to give information within the phase-space
cells with size smaller than $\epsilon$, due to the intrinsic
uncertainty exhibited by the system for finite temperatures, i.e., due
to the indistinguishability among the electronic rays. Consequently, we
would expect that $\rho_w$ violates the positivity definiteness within
some phase-space regions. 
This means that, in analogy with quantum mechanics, $\rho_w (x,p,z)$
can be defined as {\it quasi distribution}, even its $x$- and $p$-projections
are actually configuration-space  and momentum-space
distributions, respectively. 
Remarkably, from the above results it follows that it can exist a
complex function, say $\Psi (x,z)$ such that
$P_{x}(x,z)=\Psi (x,z)\Psi^{*}(x,z)$,
which is connected with $\rho_w$ by means of a
Wigner-like transform (for pure states).\cite{fgmm}
Consequently, $\Psi (x,z)$ must obey to the following
Schr\"{o}dinger-like equation:
\begin{equation}
i\epsilon(\partial \Psi /\partial z)=-
(\epsilon^2 /2)\partial ^2\Psi /\partial x^2+
U(x,z)\Psi~~~,
\label{1}
\end{equation}
which has been the starting point to construct the
quantum-like approach called the
thermal wave model (TWM).\cite{fm1}
This way, the beam as a whole
is thought as a single quantum-like particle whose {\it
diffraction-like} spreading due to the emittance (the analogous of 
$\hbar$) accounts for the {\it thermal spreading}.

Thus, we have given 
a sort of {\it Wigner-like} pictures behind
the electronic ray evolution and then recovered TWM in semiclassical 
approximation. Consequently,
solutions of (\ref{1})
for $\Psi$ in semiclassical approximation
can give solutions for the {\it deformed} equation
through Wigner transform.
It is clear that for finite emittance but in the
case in which, for $s\geq 3$,
$(\epsilon/2)^2\,\partial^2 \rho_w/
\partial p^2\gg (\epsilon /2)^s\partial^s
\rho_w/\partial p^s\quad \mbox{for}\quad s\geq 3$,
(\ref{17bis}) and its classical couterpart formally coincide for an arbitrary
(anharmonic) potential;
the similarity between $\rho$ and $\rho_w$ does not take place
for all the states.
This makes evident that for an arbitrary potential and, in particular, 
for a linear lens (harmonic oscillator) $\rho_w$ contains and $\rho$
does not contain a {\it quantum-like effect}. Worthy noting that, in 
analogy with the tomography approach in quantum mechanics and quantum 
optics,\cite{mancini-et-al-1}
we could state that in
the above quantum-like approach there is a possibility to transit from
Liouville equation to an equation for a positive marginal 
distribution,\cite{mancini-et-al-1} which has standard classical
features.

\end{document}